\documentclass{emulateapj}

\begin{document}

\shortauthors{Gordon et al.}
\shorttitle{M31: Spiral/Ring Composite Structure}

\slugcomment{ApJ Letters, in press}

\title{Spitzer/MIPS Infrared Imaging of M31: Further Evidence for 
a Spiral/Ring Composite Structure}

\author{K.~D.~Gordon\altaffilmark{1}, 
   J.~Bailin\altaffilmark{2},
   C.~W.~Engelbracht\altaffilmark{1}, 
   G.~H.~Rieke\altaffilmark{1}, 
   K.~A.~Misselt\altaffilmark{1},
   W.~B.~Latter\altaffilmark{3},
   E.~T.~Young\altaffilmark{1},
   M.~L.~N.~Ashby\altaffilmark{4},
   P.~Barmby\altaffilmark{4},
   B.~K.~Gibson\altaffilmark{2,5},
   D.~C.~Hines\altaffilmark{6},
   J.~Hinz\altaffilmark{1},
   O.~Krause\altaffilmark{1},
   D.~A.~Levine\altaffilmark{7},
   F.~R.~Marleau\altaffilmark{7},
   A.~Noriega-Crespo\altaffilmark{7},
   S.~Stolovy\altaffilmark{7},
   D.~A.~Thilker\altaffilmark{8}, and
   M.~W.~Werner\altaffilmark{9}
   }
\altaffiltext{1}{Steward Observatory, University of Arizona, Tucson, AZ 85721,
   (kgordon,cengelbracht,grieke,kmisselt,eyoung,jhinz, okrause)@as.arizona.edu}
\altaffiltext{2}{Centre for Astrophysics and Supercomputing, Swinburne 
   University of Technology, P.O. Box 218, Hawthorn, VIC 3122, Australia,
   jbailin@astro.swin.edu.au}
\altaffiltext{3}{NASA Hershel Science Center, 100-22, Caltech, Pasadena, CA 91125,
   latter@ipac.caltech.edu}
\altaffiltext{4}{Harvard-Smithsonian Center for Astrophysics, Cambridge, MA 02138,
   (mashby,pbarmby)@cfa.harvard.edu}
\altaffiltext{5}{Centre for Astrophysics, University of Central Lancashire, 
   Preston, PR1 2HE, UK, bkgibson@uclan.ac.uk}
\altaffiltext{6}{Space Science Institute, 4750 Walnut Street, Boulder, CO 80301,
   dean.hines@colorado.edu}
\altaffiltext{7}{Spitzer Science Center, 220-6, Caltech, Pasadena, CA 91125,
   (deblev,marleau,alberto,stolovy)@ipac.caltech.edu}
\altaffiltext{8}{Center for Astrophysical Sciences, Johns Hopkins Univ., Baltimore, 
   MD 21218, dthilker@pha.jhu.edu}
\altaffiltext{9}{Jet Propulsion Laboratory, MC 264-767, Pasadena, CA 91109,
   mww@ipac.caltech.edu}

\begin{abstract} 
New images of M31 at 24, 70, and 160~\micron\ taken with the Multiband
Imaging Photometer for Spitzer (MIPS) reveal the morphology of the
dust in this galaxy.  This morphology is well represented by a
composite of two logarithmic spiral arms and a circular ring (radius
$\sim$~10~kpc) of star formation offset from the nucleus.  The two
spiral arms appear to start at the ends of a bar in the nuclear region
and extend beyond the star forming ring.  As has been found in
previous work, the spiral arms are not continuous but composed of
spiral segments.  The star forming ring is very circular except
for a region near M32 where it splits.  The lack of well defined
spiral arms and the prominence of the nearly circular ring suggests
that M31 has been distorted by interactions with its satellite
galaxies.  Using new dynamical simulations of M31 interacting with M32
and NGC 205 we find that, qualitatively, such interactions can produce
an offset, split ring like that seen in the MIPS images.
\end{abstract}


\section{Introduction}
\label{sec_intro}

M31 is the nearest \citep[d~$\sim 780$~kpc,][]{Stanek98,Rich05}
external spiral galaxy \citep[Sb,][]{Hubble29} making it one of the
premier laboratories for understanding spiral galaxies like our own.
Tracing the spiral structure in M31 is an obvious first step, which
has been done with varying degrees of success for many years.
However, the high inclination (74-78$\arcdeg$) of M31 makes it
difficult to determine its large-scale structure.  In particular,
there is no clear consensus regarding the organization of spiral
structure in this important galaxy.  The earliest quantitative studies
focused on optical tracers such as HII regions
\citep{Arp63, Baade64}, OB associations \citep{Hodge79,
vandenBergh91}, and dark nebulae \citep{Hodge80}.  These studies
generally found structures consistent with a two-armed spiral
\citep[for the case for a one-armed spiral see][]{Simien78}, but
invoked disturbances such as varying inclinations and warps to account
for the deviations from simple logarithmic spirals.  More recently, HI
\citep{Guibert74, Braun91} and CO \citep{Loinard99, Nieten05} have
been used.  The velocity information available for these gas tracers
can be used to disentangle the pileup of information due to M31's high
inclination.  These studies have found strong evidence for a two-armed
spiral structure, but still suffer from the difficulty in connecting
spiral arm segments into a coherent pattern.

The difficulty in finding coherent spiral structure in M31 is
plausibly related to the interactions of M31 with its two nearest
satellites, M32 and NGC 205.  Disturbances in M31 are revealed by
tidal features far outside M31's disk visible in deep, wide-field
optical imaging \citep{Ibata01,Ibata05,Ferguson02, Lewis04}. M32 is
understood to be the perturber of the spiral arms \citep{Byrd78,
Byrd83} while NGC 205 has been modeled as the cause of the warp seen
in the optical and HI disk of M31 \citep{Sato86}.  Quantifying the
effects of M31's satellite galaxies is complicated by the lack of
measured proper motions and uncertainties in relative distances
\citep[e.g.][]{McConnachie05}.

Another difficulty in quantifying M31's spiral structure has been the
lack of appropriate tracers in the innermost regions of the galaxy.
Optical tracers are overwhelmed by the bright stellar bulge, while HI
and CO, which would be easily seen through the bulge, are largely
absent \citep{Brinks84, Melchior00} as the interstellar medium (ISM)
in the central region is composed mainly of ionized gas
\citep{Devereux94}.  The far-infrared (far-IR) emission can circumvent
these difficulties as it traces the dust in M31, which penetrates the
stellar bulge and should trace all phases of the ISM.

\begin{figure*}
\epsscale{1.05}
\plotone{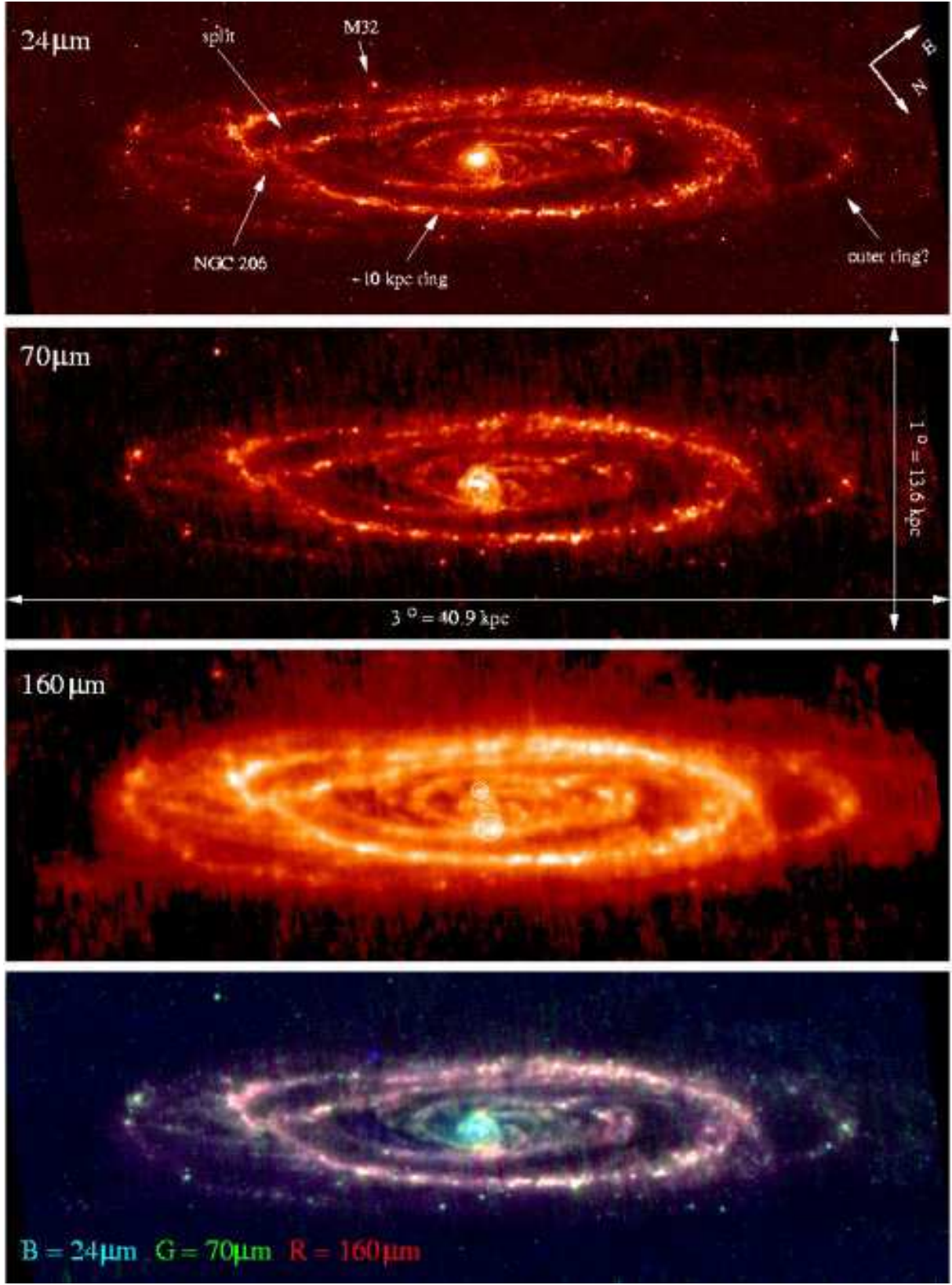}
\caption{The M31 MIPS images are shown.  The images are scaled using a
quadratic root ($x^{0.25}$) between -0.1 -- 3 MJy/sr (24~\micron), 0.1
-- 35 MJy/sr (70~\micron), and 0 -- 90 MJy/sr (160~\micron).  The
bottom image gives the 3 
color composite of the MIPS images (with scaling adjusted to emphasize
the differences between bands); the color gives an indication of
the dust temperature.  The images are oriented with the nominal
position angle of M31 ($38\arcdeg$) horizontal and are cropped to a
common size.  The nearly-vertical streaks present in the 70 and
160~\micron\ images are residual Ga:Ge detector instrumental
signatures.  M32 is clearly detected at 24~\micron, but not in the
other two MIPS bands.  NGC~205 is detected in all three MIPS bands,
but is located NW of M31 beyond the boundary of the cropped images.
The two spots near the nucleus marked on 
the 160~\micron\ image are discussed in the text.
\label{fig_m31_fullres} }
\end{figure*}

The first far-IR images of M31 were obtained with the Infrared
Astronomical Satellite (IRAS) at 12, 25, 60, and 100~\micron.  The
most prominent feature in these images is the bright, $\sim$10~kpc
radius, ring of star formation \citep{Habing84, Rice93} seen
previously in H$\alpha$ images \citep{Arp63, Devereux94}.  The
Infrared Space Observatory (ISO) imaged M31 at 175~\micron\ finding
both the bright ring and a fainter ring at larger radii
\citep{Haas98}.  The Midcourse Space Experiment (MSX) provided an
image of M31 at 8.3~\micron\ revealing possible spiral structure
inside the $\sim$10~kpc ring.  The resolutions of the IRAS ($\sim 1
\arcmin$) and ISO ($1\farcm 3$) images reveal the overall morphology,
but are not high enough to reveal the structure of the spiral arms.

We present new infrared images of M31 taken with the Multiband Imaging
Photometer for Spitzer \citep[MIPS,][]{Rieke04MIPS} instrument on the
Spitzer Space Telescope \citep[Spitzer,][]{Werner04Spitzer}.  The MIPS
spatial resolution at the distance of M31 is $6\arcsec$/23~pc,
$18\arcsec$/68~pc, and $40\arcsec$/151~pc at 24, 70, and 160~\micron,
respectively.  This paper concentrates on the newly-revealed
details of M31's infrared morphology as shown in the MIPS images

\section{Data}
\label{sec_data}

The MIPS images of M31 were taken on 25~Aug~2004.  A region
approximately $1\arcdeg \times 3\arcdeg$ oriented along the major axis
of M31 was covered with 7 scan maps.  Each map consisted of medium
rate scan legs and cross scan offsets of 148\arcsec\ with lengths
varying between $0\fdg 75$ and $1\fdg 25$ to efficiently map M31 and
its satellite galaxy NGC~205.  The MIPS DAT v2.90 \citep{Gordon05DAT}
was used to process and mosaic the individual images.  At 24~\micron\
extra steps were carried out to improve the images including readout
offset correction, array averaged background subtraction (using a low
order polynomial fit to each leg, with the region including M31
excluded from this fit), and exclusion of the 1st five images in each
scan leg due to boost frame transients.  At 70 and 160~\micron, the
extra processing step was a pixel dependent background subtraction for
each map (technique same as 24~\micron).  The background subtraction
does not remove real M31 emission as the scan legs are nearly
parallel to the minor axis with the background regions being far from
M31's disk.  The final mosaics have exposure times of $\sim$100,
$\sim$40, and $\sim$9 seconds/pixel for 24, 70, and 160~\micron,
respectively, and are shown in Figure~\ref{fig_m31_fullres}.

\begin{figure}
\epsscale{1.15}
\plotone{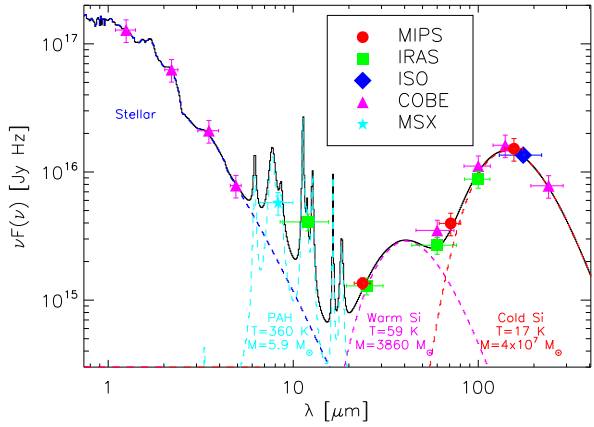}
\caption{The MIPS global fluxes of M31 are plotted along with previous
measurements from IRAS \citep{Rice88}, COBE \citep{Odenwald98}, ISO
\citep{Haas98}, and MSX \citep{Kraemer02}.  A simple stellar
population plus dust grain model (Marleau et al., in prep.) has been
fit to the data giving a 
total IR luminosity of $1.7 \times 10^{43}$~ergs~s$^{-1}$ which
corresponds to a star formation rate of 0.75~M$_{\sun}$ yr$^{-1}$
using the calibration of \citet{Kennicutt98}. \label{fig_m31_ir_sed} }
\end{figure}

The IR spectral energy distribution of M31 is shown in
Fig.~\ref{fig_m31_ir_sed}.  The global MIPS fluxes measured in a
$2\fdg 75 \times 0\fdg 75$ rectangular aperture (the background was
subtracted in the data reduction) are $107 \pm 10.7$, $940 \pm 188$,
and $7900 \pm 1580$~Jy for 24, 70, and 160~\micron, respectively.  The
uncertainties are dominated by the systematic uncertainties in the
MIPS calibration \citep{Rieke04MIPS}.  There is excellent agreement
between the MIPS and previous IR measurements.

\section{Morphology}

\begin{figure*}
\epsscale{0.95}
\plotone{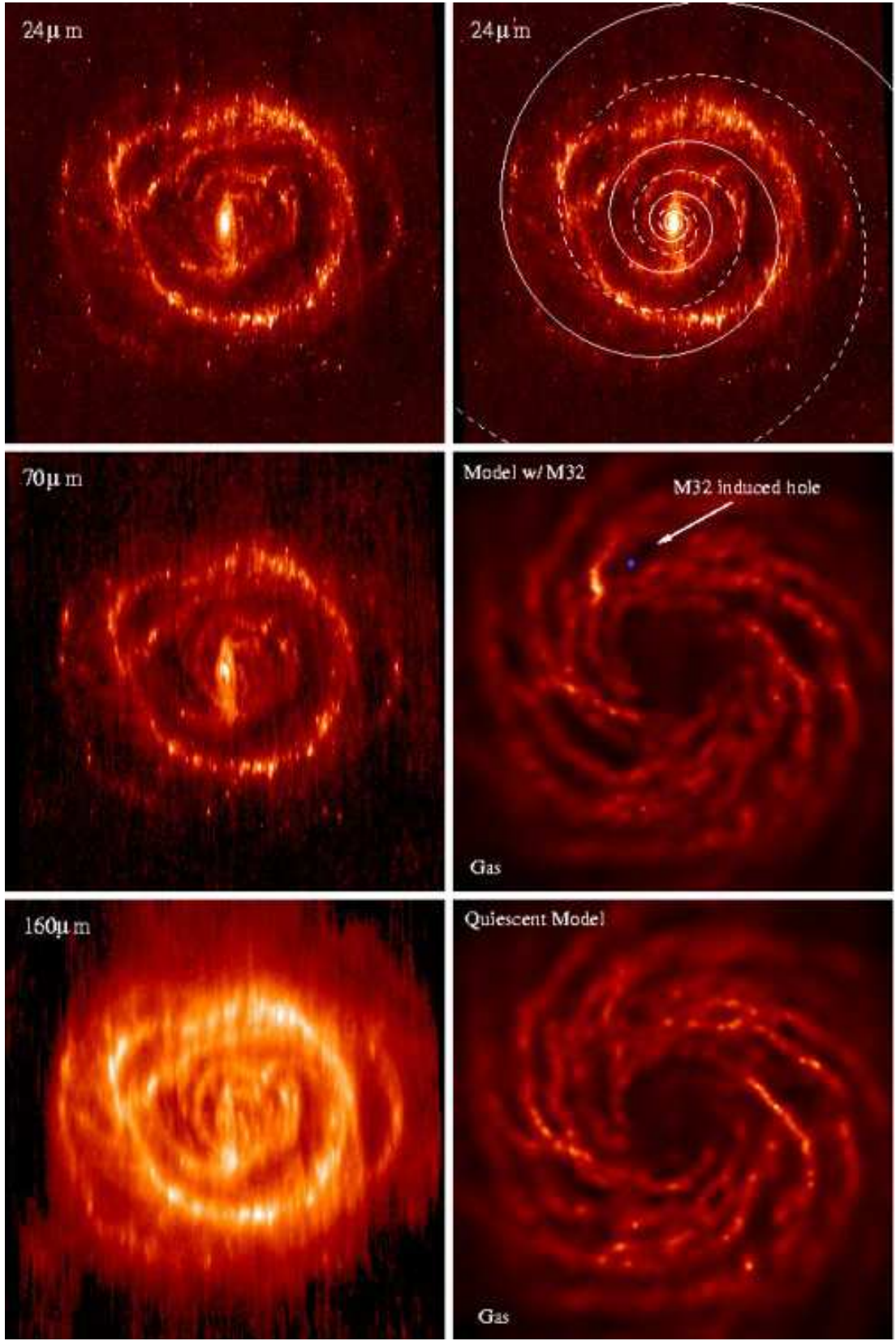}
\caption{The MIPS images deprojected using an inclination of
$75\arcdeg$ are shown in the left column.  The right column shows the
24~\micron\ image with the spiral arms shown in
Fig.~\ref{fig_m31_ir_regions} overplotted, the M31-M32 interaction
model (M32 as a blue dot), and M31 quiescent model (see
\S\ref{sec_model}).  The M31-M32 interaction model image gives the
true location of M32.  If deprojected with the same assumptions as the
images, the location of M32 would shift off the image (within the
$\sim$10$\arcmin$ model insensitivity to the true position of M32).
The model images give the integrated gas density which is nominally
proportional to the dust given a constant gas-to-dust ratio (except in
the inner region where it is clear that the gas-to-dust ratio likely
varies significantly).  The orientation of the images is the same as
in Fig.~\ref{fig_m31_fullres} and the size is $3\arcdeg \times
3\arcdeg$.
\label{fig_m31_deproj} }
\end{figure*}

The MIPS images in Fig.~\ref{fig_m31_fullres} are dominated by the
$\sim$10~kpc ring.  The deprojected ISO image showed that the ring is
fairly circular, but with an obvious splitting near the position
of M32 \citep{Haas98}.  This splitting is now seen to extend over
$\sim$1/4 of the ring.  The outer ring seen with ISO is seen in the
MIPS images, but appears to be relatively incomplete.  Overall, the
morphology of the disk of M31 in the infrared appears more disturbed
on the left-hand (southwest) than right-hand (northeast) side of the
galaxy.

The MIPS images show emission extending beyond their edges along the
major axis, implying that the infrared extent of M31 is larger than
$3\arcdeg$/40.9~kpc.  A comparison of the MIPS 160~\micron\ image with the
latest HI image (Braun et al., in prep.) reveals dust emission where
there is HI emission.  The gas and dust seem well mixed even in the
outer regions of M31.

The infrared emission in the nuclear region has a component with
hotter dust than the rest of M31 (but also a cold component due to the
emission seen in the 160~\micron\ image).  This is consistent with the
nuclear region gas being dominated by ionized gas as seen in H$\alpha$
\citep{Devereux94} as well as emission from AGB stars in the bulge.
As the flux from the nucleus fades towards longer wavelengths, the two
spots above and below along the minor axis gain prominence.  The
prominence of the spot above the nucleus was noticed by \citet{Rice93}
and confirmed by \citet{Haas98}.  These spots appear to be the ends of
a bar and the beginning of spiral arms.  The location of these spots
and appearance of the nuclear region (especially at 70~\micron) is
similar to the triaxial bulge simulations of \citet{Berman01}.

To investigate the morphology in more detail, the images have been
deprojected assuming an inclination of $75\arcdeg$, and displayed in
Fig.~\ref{fig_m31_deproj}.  Usually the inclination is picked visually
to make the galaxy most spiral-like or circular in the deprojected
frame.  To be more objective, the inclination we use is the median
value of the independently calculated inclinations of the spiral arm
segments investigated by \citet{Braun91}.

Arc-like structures dominate the morphology of the region inside the
$\sim$10~kpc ring and appear to be spiral arms with starting points at
ends of a bar.  The existence of a bar in M31 has been
suggested in the past \citep{Stark94}.  Since circularly symmetric
objects will deproject into elongated bars, the evidence for a bar is
based on the prominence of the two spots near the nucleus at
160~\micron, not on the appearance of a bar in the nuclear region at
24~\micron.

\begin{figure*}
\epsscale{1.15}
\plotone{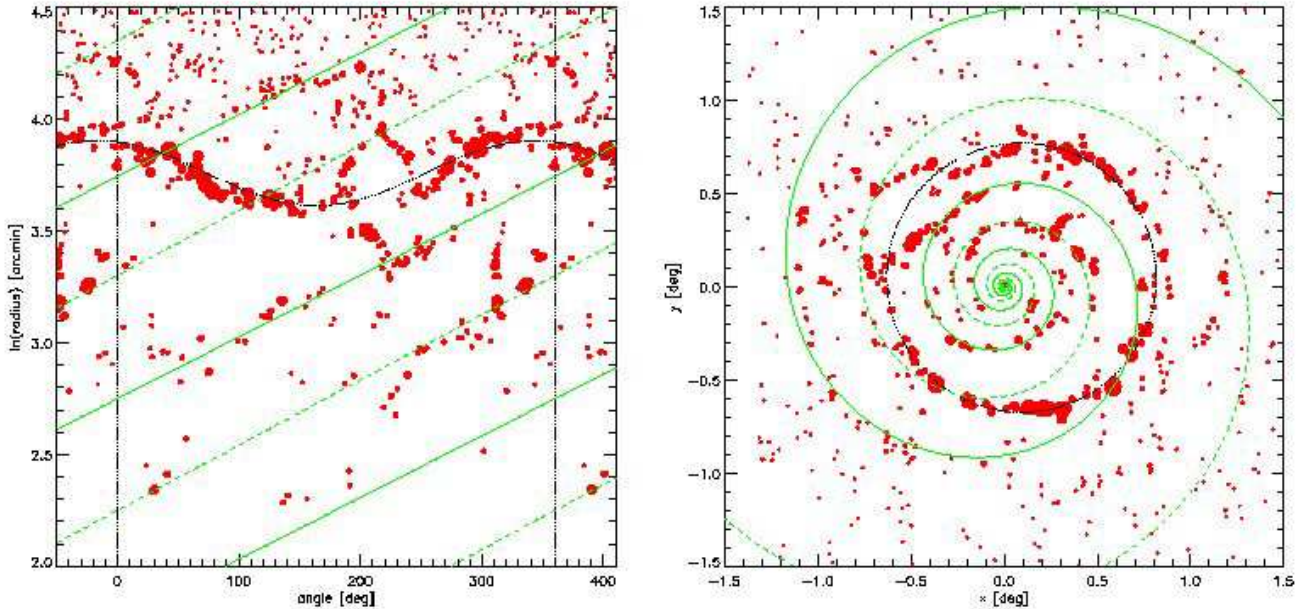}
\caption{Polar and Cartesian plots showing the locations of point
sources detected in all three MIPS bands at 160~\micron\ resolution.
The symbols have been scaled by the measured 160~\micron\ fluxes.
These plots are overlaid with two simple logarithmic spirals (green,
one solid and one dashed) and an offset circle (black).
\label{fig_m31_ir_regions} }
\end{figure*}

To determine the parameters of any spiral arms,
Fig.~\ref{fig_m31_ir_regions} gives plots in polar and Cartesian
coordinates of the locations of compact IR objects seen in all three
MIPS bands.  These regions are likely to be star forming regions
(e.g., HII regions and OB associations) which are excellent spiral arm
tracers.  Point source photometry was done using StarFinder
\citep{Diolaiti00} after convolving the 24 and 70~\micron\ images to
the 160~\micron\ resolution using custom convolution kernels (Gordon
et al., in prep.) ensuring the same physical regions were measured in
the three MIPS bands.

In polar coordinates, a logarithmic spiral is a line with a non-zero
slope and a circle is a horizontal line.  The strongest feature is the
wavy line around ln(radius) = 3.75 representing the $\sim$10~kpc ring.
This wavy line is fit well by a circle offset from the nucleus by
(5.5\arcmin, 3.0\arcmin) with a radius of 43\arcmin/9.8~kpc, except
for the region around $200\arcdeg$ where the ring splits.  The
remarkable circularity of this ring makes it difficult to explain this
structure as spiral arm segments.  Two logarithmic spirals are plotted
which roughly follow the spiral structure.  These trailing spirals
have pitch angles of 9$\arcdeg$ and $9\fdg 5$, and are phase shifted by
166$\arcdeg$.  These two spirals trace most of the structure inside
the ring as well as some of the structure outside the ring.  These two
spirals are not unique solutions, but reasonable given the disturbed
nature of M31.  The pitch angles used here are larger than found in
previous work, consistent with not including the offset $\sim$10~kpc
ring as part of the spiral structure.  We found that our simple
logarithmic spirals plus an offset circle give as good or better
representation of the MIPS sources than the \citet{Braun91} spiral arm
segments when the segments were deprojected in the same manner as the
MIPS sources.

Finding that M31 has an offset, star forming ring which is not
just a superposition of spiral arm segments raises the question:
``What is the origin of the ring?''  The ring could be due to internal
resonances which can cause three different types of rings: nuclear,
inner, and outer \citep{Buta86}.  The M31 ring is clearly not nuclear
or an inner ring which is directly associated with bar.  The M31 ring
could be an outer ring which is associated with the outer Lindblad
resonance/radius of corotation.  The radius of corotation is placed
around 18~kpc by \citet{Braun91}.  The ring could also be caused by
interactions like those seen in the Cartwheel galaxy
\citep{Hernquist93}.  An interesting clue to the possible origin of
the ring is the splitting of the ring by M32's passage through M31's
disk ($\S$\ref{sec_model}).  Is this a coincidence or cause of the
ring?  A detailed study of the origin of M31's ring is clearly needed.

While the logarithmic spirals plotted in Fig.~\ref{fig_m31_ir_regions}
represent the data well, there are clear deviations from these spirals
in addition to the well defined circle of star formation.  The known
interactions between M31 and its satellite galaxies M32 and NGC 205
imply these interactions may be the cause of the deviations
from regular spiral structures.

\section{Dynamical Models}
\label{sec_model}

The dynamical effects that M32 and NGC 205 have on M31 were
investigated by running a number of numerical simulations of the
M31-M32 and M31-NGC 205 interactions using GADGET2 \citep{Springel01,
Springel05}.  These simulations self-consistently model gravity and
hydrodynamics.  In addition, we have implemented star formation
following \citet{Katz96}, cooling using the MAPPINGS
III (http://www.mso.anu.edu.au/$\sim$ralph/map.html)
collisional ionization equilibrium models for [Fe/H]=-0.5, and a
combination of thermal and kinetic feedback from supernovae
\citep[with $f_v=0.1$]{Navarro93} using the instantaneous
recycling approximation.  We used only the tree algorithm when
calculating the gravitational forces as we found that the default
TreePM method introduced unacceptably large numerical artifacts.

The initial conditions for M31 were the ``M31a'' equilibrium
disk-bulge-halo model of \citet{Widrow05}, with 132092, 48000, and
357192 particles in each component respectively.  A randomly-sampled
fraction of the resulting disk particles were turned into gas
particles such that the radial gas density profile followed the sum of
the HI (Braun et al., in prep.) and CO \citep{Nieten05} gas profiles
except in the inner region where we assume a constant density to
account for the mass of ionized gas.  M32 and NGC~205 are modeled as
point particles, with radial velocities taken from \citet{Mateo98}.
The current line-of-sight distances to the satellite galaxies were
varied within the uncertainties of current estimates (M32:
\citet{McConnachie05}; NGC~205: \citet{Ferrarese00}), and the
velocities in the plane of the sky were varied freely in order to find
configurations where each satellite has recently passed through the
disk of M31.  

In Figure~\ref{fig_m31_deproj} we show an image of one such simulation
of M31 after an interaction with M32, assuming that M32 currently lies
8~kpc behind the nucleus of M31, has a velocity in the plane of the
sky of $200~\mathrm{km~s^{-1}}$ SSE, and has a mass of $1 \times
10^{10}~\mathrm{M_{\Sun}}$ (3--5 times larger than current estimates
\citep{Mateo98}).  We also show an image of a simulation with no
satellite.  The passage of M32 through the disk of M31 results in a
burst of star formation that propagates outward through the disk
resulting in a large hole similar in size and location to the observed
splitting of the star forming ring.  The location and age of the star
forming region NGC~206 is consistent with this star formation
\citep{Hunter96}.  We find that such a feature rapidly distorts due to
differential rotation, and so its present symmetric appearance
indicates that the disk passage occurred very recently (20~Myr ago in
this model).  If we use a mass of $2 \times 10^{9}~\mathrm{M_{\Sun}}$
for M32, as suggested by \citet{Mateo98}, very little effect is seen
on the disk of M31, suggesting that M32 may be more massive than
previously thought.  It is interesting to note that the input high
density gas ring is pulled off-center ($\sim$$1\arcmin$) by the
satellite interaction, similar but smaller than the observed offset
($6.3\arcmin$).  The simulated ring fragments into spiral arms and,
therefore, is not as apparent as the observed ring.

The mass of NGC~205 is uncertain (\citet{Mateo98} calculates a mass of
approximately $8 \times 10^{8}~\mathrm{M_{\Sun}}$) and therefore its
ability to strongly affect the disk of M31 is equally uncertain.  We
have run a number of simulations of NGC~205, and while we can find
orbits for which it passes through and temporarily distorts the spiral
structure in the disk, we have been unable to find one which produces
a hole that matches the observed split in the ring.  However, an
exhaustive search of the parameter space has not yet been carried out.

It is a generic feature in all of our models (with or without
satellite interaction) that the gas in the central region of M31 ends
up at a temperature greater than $10^4$~K, consistent with the
observation of ionized gas in the nuclear region.
This behavior is not seen in models with a fully exponential gas disk;
therefore, the hot nucleus is a direct result of the central
depression of the gas density.  As central depressions are commonly
seen in the HI profiles of disk galaxies \citep[e.g.][]{Hewitt83}, our
simulations suggest that a large fraction of the gas in the centers of
disk galaxies is ionized.

\section{Conclusions}

We have presented new MIPS images of M31 at 24, 70, and 160~\micron\
which have higher spatial resolution and sensitivity
than previous infrared images.  These new images reveal the presence
of two spiral arms in addition to a ring of star formation.  The
spiral arms appear to emanate from a bar in the nuclear region and can
be roughly traced out beyond the $\sim$10~kpc ring of star formation.
The ring of star formation forms an almost complete circle which is
offset from the nucleus and split near the location of M32.  This ring
of star formation and the disturbed spiral arms are indications that
interactions may be affecting the morphology of M31.  Dynamical models
of M31 interacting with M32 predict morphology qualitatively similar
to M31's (while models with NGC~205 instead of M32 give only small
perturbations).  Specifically, both the splitting and the offset of
the star forming ring are seen as well as copious spiral arms.  It
seems that M31 is not an undisturbed normal spiral galaxy, but
one which has been significantly affected by interactions.

This paper presents a preview of the possibilities becoming available
to understand M31.  Data covering the majority of this galaxy are
available in the X-ray (XMM, Trudolyubov et al., in prep.),
ultraviolet \citep[GALEX,][]{Thilker05}, optical (Engelbracht et al.,
in prep.), mid-infrared (Spitzer/IRAC, Barmby et al., in prep.),
infrared (this paper), HI (Braun et al., in prep.), CO
\citep{Nieten05}, and radio continuum \citep{Beck89}; these will
enable us to make detailed models on the impact that satellite
galaxies have on the morphology of M31. 

\acknowledgements
This work is based on observations made with the {\em Spitzer Space
Telescope}, which is operated by the Jet Propulsion Laboratory,
California Institute of Technology under NASA contract 1407. Support
for this work was provided by NASA through Contract Number \#1255094
issued by JPL/Caltech.  JB thanks Paul Bourke for his visualization
work and John Dubinski for providing an updated copy of the GalactICs
code.

\end{document}